\documentclass[12pt]{article}
\usepackage{amssymb}\usepackage{bm}\usepackage{amsfonts}\usepackage{amsmath}\usepackage{hyperref}
\newcommand{\beq}{\begin{equation}}\newcommand{\eeq}{\end{equation}}\newcommand{\beqa}{\begin{eqnarray}}
\newcommand{\eeqa}{\end{eqnarray}}\newcommand{\ts}{\textstyle}
\newcommand{\dl}{\bm{\delta}}\newcommand{\nn}{\nonumber}

\newcommand{\Lie}[1]{\mathcal{L}_{\bar{#1}}}
\newcommand{\GLie}[1]{\widetilde{\mathcal{L}}_{\bar{#1}}}
\newcommand{\h}[1]{\hat{#1}}
\newcommand{\ulambda}{\underline{\lambda}}

\def\be{\begin{equation}}
\def\ee{\end{equation}}
\def\ba{\begin{eqnarray}}
\def\ea{\end{eqnarray}}

\def\dd{\textrm{d}}

\def\f{\frac}
\def\p{\partial}
\def\h{\hat}

\begin{document}
{\renewcommand{\thefootnote}{\fnsymbol{footnote}}
\hfill  Phys. Rev. D {\bf 80}, 044006 (2009)\quad  \quad \quad IGC--05/5--4\\
\medskip
\begin{center}
{\LARGE  The Internal Spin Angular Momentum of an Asymptotically Flat Spacetime}\\
\vspace{1.5em}
Andrew Randono\footnote{e-mail address: {\tt arandono@gravity.psu.edu}}, David Sloan\footnote{e-mail address: {\tt sloan@gravity.psu.edu}}
\\
\vspace{0.5em}
Institute for Gravitation and the Cosmos,\\
The Pennsylvania State
University,\\
104 Davey Lab, University Park, PA 16802, USA\\
\vspace{1.5em}
\end{center}
}

\setcounter{footnote}{0}
\begin{abstract}
In this paper we investigate the manner in which the internal spin angular momentum of a spinor field is encoded in the gravitational field at asymptotic infinity. The inclusion of internal spin requires us to re-analyze our notion of asymptotic flatness. In particular, the Poincar\'{e} symmetry at asymptotic infinity must replaced by a spin-enlarged Poincar\'{e} symmetry. Likewise, the generators of the asymptotic symmetry group must be supplemented to account for the internal spin. In the Hamiltonian framework of first order Einstein-Cartan gravity, the extra generator comes from the boundary term of the Gauss constraint in the asymptotically flat context. With the additional term, we establish the relations among the Noether charges of a Dirac field, the Komar integral, and the asymptotic ADM-like geometric integral. We show that by imposing mild restraints on the generating functionals of gauge transformations at asymptotic infinity, the phase space is rendered explicitly finite. We construct the energy-momentum and the new total (spin+orbital) angular momentum boundary integrals that satisfy the appropriate algebra to be the generators of the spin-enlarged Poincar\'{e} symmetry. This demonstrates that the internal spin is encoded in the tetrad at asymptotic infinity. In addition, we find that a new conserved and (spin-enlarged) Poincar\'{e} invariant charge emerges that is associated with the global structure of a gauge transformation. 
\end{abstract}

\section{Introduction}
\textit{What happens when a spinor falls into a black hole?} More specifically, what happens to the \textit{spin} of the spinor when it falls into a black hole? For simplicity let's assume the blackhole is initially uncharged and non-rotating. Suppose the spinor has fallen behind the horizon, so the spinor field has support only in a region inside the horizon and there is no way that an observer at constant radius outside the black hole can extract any information directly from measurements of the the field itself. What information about the spinor field can an observer at rest at spatial infinity extract from classical measurments of the available fields at infinity\footnote{Throughout this paper we will ignore the possibility of information leakage through quantum effects like Hawking evaporation.}. Clearly the mass of the spinor is information that is available to the observer since the mass of the black hole will increase, and the ADM mass is a quantity which one can compute from the gravitational field at asymptotic infinity. Similarly, the ordinary angular momentum of the black hole may increase dues to the addition of the orbital angular momentum of the spinor field, and this is encoded in the asymptotic gravitational field via a similar ADM expression. The charge of the spinor, since it couples to a long-range force, can be extracted through measurements of the electromagnetic field at spatial infinity. What about the internal spin angular momentum? One may argue that the internal spin also couples to the electromagnetic field giving rise to a magnetic moment, whose value is a unique characteristic of spinors. \textit{But, what if the spinor is a neutrino?} The neutrino does not couple to any long range forces other than gravity. Any information about the field from its coupling to the electroweak force will be well hidden behind the horizon. Can the internal, half-integer spin be encoded somehow in the gravitational field at asymptotic infinity? At first glance this may seem peculiar since half integer spin is a property unique to fermionic matter, and the gravitational field is generally strictly associated with bosonic properties.

In this paper we argue that the internal spin can indeed be encoded in the gravitational field of an asymptotically flat spacetime. We will work entirely in the first order, Einstein-Cartan framework of gravity, where the torsion is allowed to be non-zero. Coupling the gravitational field to spinors, we will investigate the total angular momentum of an asymptotically flat spacetime. Working with the first order framework has the advantage that the full theory is finite without the addition of infinite counter-terms as has been recently shown \cite{AshtekarSloan,AshtekarSloan2}. Here we will show that in the first order framework, the total angular momentum picks up additional terms in order to account for the internal spin of the fermionic matter in the interior of the spacetime.

In an open universe, typically each constraint of the total Hamiltonian must pick up a boundary term at asymptotic infinity to make the theory finite. These boundary terms turn the total Hamiltonian constraint into a true, non-zero Hamiltonian and each component is the generator of the Poincar\'{e} symmetry of the spacetime at asymptotic infinity. In addition to the Hamiltonian constraint and the diffeomorphism constraint whose boundary terms generate asymptotic translations (or proper Lorentz transformations with a different choice for the Killing vector), the first order framework has an additional constraint, the Gauss constraint, which generates local Lorentz transformations of objects living in the $Spin(3,1)$ representation space, like spinors. 

Although expressions for the total angular momentum of asymptotically flat spacetimes have been considered in the past including spinors coupled to the gravitational field via the tetrad, such investigations typically relied on the second order framework where the connection is expressed as an explicit function of derivatives of the tetrad. In this case, since the tetrad is tensorial it has been claimed that the Gauss constraint does not pick up a boundary term and there is no additional term in the angular momentum to account for internal spin \cite{Teitelboim:ECSurface,Teitelboim:Fermions,Teitelboim:Supergravity}.  Here we show that in a true first order framework the Gauss constraint must pick up a boundary term which we identify with the generator of internal Lorentz transformations of the total internal spin angular momentum of the spacetime. As usual, the new \textit{bulk} + \textit{boundary} term is generically non-zero, and it can be evaluated to give the total internal spin angular momentum which includes the new contribution to account for the internal spin of fermions in the interior.

The physical mechanism that allows for this identification of the new boundary term with the internal spin is the coupling of the internal spin of the spinor field with the gravitational field via the tetrad. Working in the first order framework, thereby allowing for torsion, will allow us to exploit recent results concerning the finiteness of asymptotically flat first order gravity \cite{AshtekarSloan, AshtekarSloan2}. In the standard, minimal first-order framework, the torsion does not self-propagate and it is non-zero only when the axial current is non-zero. Since we will assume the spinor fields fall off sufficiently rapidly outside a region of compact to be negligible at the boundary, the torsion is effectively zero at asymptotic infinity. Since the results we will derive are only strictly known to be valid in the first order framework, the precise role of torsion remains unclear. For example, it is not known if the existence of non-zero torsion in the interior affects the geometry near asymptotic infinity to encode the internal spin, or if such an effect is a more generic property of the coupling of spinors to the tetrad which would remain even in a torsion free, second order framework. As mentioned previously, some evidence that the latter scenario is not the case has been presented in the literature \cite{Teitelboim:ECSurface,Teitelboim:Fermions,Teitelboim:Supergravity}, however, a full analysis of the second order theory is necessary to reach a definite conclusion on this matter.

\subsection{Conventions} Our conventions used in this paper are as follows. We assume that spacetime, $(M,e)$ is a four-dimensional Riemannian manifold with metric signature $(-,+,+,+)$. We will work with Grassman-valued spinor fields, which we will assume to be Dirac spinors. To ensure that the inner product on Dirac spinors is real, this signature requires that we define the dual spinor with an extra factor of $i$ so that $\bar{\psi}=i\,\psi^{\dagger}\gamma^0$. Generically upper case Roman characters $\{I,J,K,...\}$ will refer to internal Lorentz indices, while lower case Greek indices $\{\alpha,\beta,\mu,\nu,...\}$ will refer to spacetime indices. Since we will be dealing with asymptotically flat spacetimes, we will have occasion to introduce an abstract vector space whose elements transform under a vector representation of the Lorentz group. Objects in this vector space will be denoted by hatted uppercase Roman indices $\{\h{I},\h{J},\h{K},...\}$. Usually we will work in an index free Clifford algebra notation where the tetrad is valued in the vector elements of the Clifford algebra, $e\equiv \frac{1}{2} \gamma_I e^I$, and the spin connection, $\omega\equiv \frac{1}{4} \gamma_I \gamma_J \,\omega^{[IJ]}$, takes values in the bivector elements of the algebra. The spin connection is not assumed to be torsion free, so its torsion and curvature are given by
 $T=D_\omega e=\frac{1}{2}\gamma_I T^I$, and $R_\omega=d\omega+\omega\wedge \omega =\frac{1}{4}\gamma_I \gamma_J \,R^{IJ}$. The duality operator is given by $\star =-i\gamma_5=\gamma^0\gamma^1\gamma^2\gamma^3=\frac{1}{4!}\epsilon_{IJKL}\gamma^I\gamma^J\gamma^K\gamma^L$ with $\epsilon_{0123}=-\epsilon^{0123}=1$. The upppercase $D$ will be reserved for the exterior covariant derivative with respect to the spin-connection: $D=D_\omega=d+\omega$. To avoid proliferation of symbols, we will drop explicit wedge products between differential forms, and in integral expressions the trace over the Clifford algebra in a $4\times 4$ matrix representation is assumed. To distinguish vectors and forms on $M$ from the vectors and forms defined on the infinite dimensional phase space, we will denote the latter in \textbf{bold} font. We will use two notations for the contraction operation denoted $\alpha(\bar{V})=\iota_{\bar{V}}\alpha=\alpha(\bar{V}, \,...\,, \ )$.
 
\subsection{Outline} 
The outline for this paper is as follows. Section (\ref{sec:Spin}) begins with a basic review of angular momentum including the internal spin angular momentum of fundamental spinor fields. Our review focuses on the algebraic properties of the spin enlarged Poincar\'{e} symmetry that establish the notion of conserved total angular momentum. In section (\ref{sec:Set-up}), we outline the basic set-up for the problem, detailing the first order Einstein-Cartan action with an asymptotic boundary and fermionic matter, and the construction of the symplectic form in the covariant phase space approach. We show that in Einstein-Cartan gravity, in addition to the Hamiltonian and momentum constraints, the Gauss constraint also must be supplemented by a boundary term which generates internal $Spin(3,1)$ transformations on the asymptotic boundary. We then review, in section (\ref{sec:Noether}), the construction of the Noether charges corresponding to energy-momentum and total (spin+orbital) angular momentum of the Dirac field focusing on some subtleties of the group structure that must carry over into the asymptotic symmetry group. We conclude this section with a generalization of the Noether charges to spacetimes with curvature and torsion. In section (\ref{sec:Komar}), by assuming the existence of a global Killing vector, we use the previously constructed symplectic form and the generators of the spin-enlarged Poincar\'{e} symmetry to reveal the relations between the geometric boundary integrals of gravitational fields and the bulk integrals of matter fields. Here we highlight the role of the additional boundary term coming from the Gauss constraint and its relation to the internal spin angular momentum. This additional term allows us to establish the relations among the Noether charges, the Komar integral, and an ADM-like asymptotic integral. In section (\ref{sec:Asymptotics}), we lift the restriction on the existence of a global Killing vector and assume that we only have asymptotic (spin-enlarged) Poincar\'{e} symmetry. Using recent results on the finiteness of first order Einstein-Cartan gravity, we establish that the additional Gauss surface term is explicitly finite as well so long as we impose mild restrictions on the set of allowed gauge transformations near asymptotic infinity. Continuing with this analysis in section (\ref{sec:Generators}), we show that requiring that the local gauge group of Einstein-Cartan gravity reduces to the spin-enlarged Poincar\'{e} group at asymptotic infinity demands further restrictions on the allowed internal gauge transformations at asymptotic infinity. We show that with these restrictions, the energy-momentum and total (spin+orbital) angular momentum geometric boundary integrals are gauge invariant and satisfy the spin-enlarged Poincar\'{e} algebra under the Poisson bracket, and we construct the two invariants from the Casimirs of the algebra. We show that, surprisingly, a new (spin-enlarged) Poincar\'{e} invariant charge emerges from the construction that appears to be associated with global information contained in a gauge transformation itself.

\section{Spin \label{sec:Spin}}
Here we will recall some basic facts about internal spin angular momentum. We wish to emphasize that the existence of conserved internal angular momentum follows from pure algebraic considerations alone, apart from the dynamics of the particle. 

Recall that Poincar\'{e} algebra can be obtained from a Wigner-In\"{o}n\"{u} contraction of the de Sitter or anti-de Sitter algebra by taking the limit as the ``cosmological constant'' goes to zero \cite{Ortin}. The (A)dS algebra is formed by the Lorentz generators $\lambda=\frac{1}{2}\lambda_{IJ}J^{IJ}$ together with the pseudo-translation generators, here scaled by the cosmological parameter $\ell$. The (A)dS algebra is then
\beqa
\left[ \lambda_1 , \lambda_2\right] &=& {\lambda^I_{1}}_K \,\lambda^{KJ}_{2}\,{\ts \frac{1}{2}}J_{IJ} \nn\\
\left[ \lambda, \eta \right] &=& \lambda_{IK}\eta^K\,{\ts \frac{1}{\ell}}P^I \nn\\
\left[ \eta_1,\eta_2 \right]&=& {\ts \frac{1}{\ell^2}} \,\eta^{[I}_{1}\eta^{J]}_{2}\,{\ts \frac{1}{2}}J_{IJ}\,.
\eeqa
The Poincar\'{e} algebra is obtained by taking the limit as $\ell \rightarrow 0$, keeping terms linear in $\ell$ and dropping higher order terms. The algebra is unchanged apart from the commutator of two translations, which now commute. Being second rank, semi-simple algebras, the de Sitter algebra, $dS_4\simeq SO(4,1)$, and the anti-de Sitter algebra $AdS_4\simeq SO(3,2)$ both possess two invariant Casimir operators. The standard quadratic Casimir contracts to the conserved mass of the Poincar\'{e} algebra:
\beqa
C_2= P_I P^I\,.
\eeqa
In addition to this, the algebras possess a quadratic Casimir, which also contracts to an operator on the Poincar\'{e} algebra, given by $C_4=W_I W^I$ where $W_I=\frac{1}{2} \epsilon_{IJKL}J^{JK}P^L$. In the massive case (which due to the positive energy theorem will be the case of primary concern) this operator can be re-expressed as
\beq
C_4=-C_2 \,{\ts \frac{1}{2}}\tilde{J}_{IJ}\tilde{J}^{IJ}
\eeq
where $\tilde{J}$ is the perpendicular projection of $\sigma$ into the components perpendicular to the momentum:
\beq
\tilde{J}^{IJ}=\perp^I_K \perp^J_L \,J^{KL}\quad\quad\quad \perp^I_J=\delta^I_J-P^I P_J /P\cdot P \,.
\eeq
The generic eigenvalues of the quadratic Casimir take on continuous values (the mass squared), $C_2=-M^2$. On the other hand, the quotient $C_4/C_2$ takes on discrete values in an irreducible representation yielding the relativistic angular momentum invariant: 
\beq
C_4/C_2 =s(s+1)\quad\quad \textstyle{with}\quad \quad s\in \mathbb{Z}/2\,.
\eeq
In flat spacetime, the single particle, spinor realization of the Poincar\'{e} algebra consists of the translations, which are the ordinary momentum operators $P_\mu=-i \partial_\mu$, and the angular momentum generators given by $J^{\alpha\beta}=j^{\alpha\beta}_{orbital}+j^{\alpha\beta}_{spin}$. Here $j^{\alpha\beta}_{orbital}=-i(x^\alpha \partial^\beta -x^\beta \partial^\alpha)$ is the (relativistic) orbital angular momentum operator, and $j^{\alpha\beta}_{spin}=\sigma^{\alpha\beta}=\frac{1}{2}\gamma^{[\alpha}\gamma^{\beta]}$ is the genrator of internal Lorentz transformations. In the single particle case, only the spin angular momentum contributes:
\beq
C_4/C_2=-{\ts \frac{1}{2}}\tilde{\sigma}_{IJ}\tilde{\sigma}^{IJ}={\ts \frac{1}{2}}\left({\ts\frac{1}{2}}+1\right)\,.
\eeq
However, generically in a multi-particle theory or field theory, both the orbital and spin angular momentum components will contribute to the relativistic angular momentum invariant. 

Since the spin invariant is constructed from purely algebraic considerations, the notion of internal spin should be inherent in any realization of the Poincar\'{e} algebra. In particular, the isometries of an asymptotically flat spacetime spacetime form a realization of the Poincar\'{e} algebra. Since the Einstein equations coupling gravity to spinors admit asymptotically flat solutions, the total spin of the matter content in the interior of the spacetime should be encoded in the geometry at asymptotic infinity, and the total invariant spin of the spacetime should be encoded in the generators of the realization of the Poincar\'{e} algebra. However, to our knowledge this has not yet been realized in the literature.

From the explicit form of $C_4/C_2$, it is clear that preserving the full covariance of the Poincar\'{e} algebra is essential in the construction. One might expect that since $C_4/C_2$ generically has the same eigenvalues as the Casimir of $SU(2)$, a gauge fixing which preserves the $SU(2)$ subgroup may still contain all the necessary information about the spin to construct the spin invariant. More explicitly, given a particular value for the momentum $P^I$, one can always perform a group reduction from a faithful representation of $Spin(3,1)$ to a faithful representation of only the subgroup $SU_{(P)}(2)$ that preserves the momentum 4-vector $P^I$. On this restricted set with fixed momentum, the Casimir $C_4/C_2$ is identical to the Casimir of the representation of $SU_{(P)}(2)$, which explains why the Casimir takes on the discrete spectrum characteristic of non-relativistic angular momentum. On the other hand, the momentum vector is not given \textit{a priori} --- it will be different for each realization. Thus, to fully capture the internal spin angular momentum, one must work in a completely covariant framework where the full Poincar\'{e} algebra is preserved. Any gauge fixing will eliminate physical degrees of freedom. For our purposes, this means that one must work with a completely covariant framework of the Hamiltonian analysis of Einstein-Cartan gravity \cite{Alexandrov:Covariant,Randono:CanonicalLagrangian,Randono:CovariantCanonical} as opposed to, for example, the common gauge choice of the time gauge where the gauge group is reduced by hand by fixing a timelike vector field in the fiber. Throughout this paper we will adopt the ``covariant phase space" approach \cite{Carlip:Book, AshtekarMagnon, CrnkovicWitten, Crnkovic} where the phase space is restricted to be the set of solutions to the Einstein-Cartan equations of motion. This allows us to define a conserved symplectic form without the need for gauge fixing. 

\section{The Set-up \label{sec:Set-up}}
Consider the action of first order, Einstein-Cartan gravity coupled to a spinor field:
\beqa
S&=& S_{grav}+S_{ferm}
\eeqa
with the gravitational part of the action given by the Einstein-Cartan action together with a surface term at asymptotic infinity:
\beqa
S_{grav}&=&\frac{1}{k}\int_{M}\star \,e\,e\,R-\frac{1}{k}\int_{\partial M}\star \,e\,e\,\omega\nn\\
&=&\frac{1}{4k}\int_M \epsilon_{IJKL}\,e^I\,e^J\,R^{KL} -\frac{1}{4k}\int_{\partial M}\epsilon_{IJKL}\, e^I\,e^J\,\omega^{KL}\,.
\eeqa
Here the spacetime manifold $M$ is a taken to be a differential manifold with a timelike boundary at infinity, $\partial M=\tau_\infty$. As usual in the canonical framework, we also assume the manifold has the topological structure $M=\Sigma\times \mathbb{R}$, where $\Sigma$ is a spatial hypersurface with $\partial \Sigma=S^2$ is the boundary at spatial infinity. The surface term is chosen so that when the phase space corresponding to an asymptotically flat spacetime is chosen carefully, the gravitational action is functionally differentiable \cite{AshtekarSloan,AshtekarSloan2}. 

The fermionic part of the action we will take to be a massive Dirac field with either compact support or fall-off conditions such that boundary terms can be ignored, whose action is 
\beqa
S_{ferm}&=&\alpha \int_M\frac{1}{2}\left(\overline{\psi}\star e\,e\,e\,D\psi+\overline{D\psi}\star e\,e\,e \,\psi \right) -m \bar{\psi}{\psi}\star e\,e\,e\,e \nn\\
&=& \alpha \int_M \frac{1}{2} \epsilon_{IJKL}e^I\,e^J\,e^K \left( \overline{\psi}\gamma^L\,D\psi-\overline{D\psi}\gamma^L \psi \right)-\frac{m}{4} \bar{\psi}\psi\,\epsilon_{IJKL}e^I\,e^J\,e^K\,e^L \nn\\
&=& 6\alpha \int_M d^4x\,e\left( \frac{1}{2} e^\mu_I (\bar{\psi}\gamma^I D_\mu \psi-D_\mu \bar{\psi}\gamma^I \psi)- m\bar{\psi}\psi\right)\,.
\eeqa

The variation of the action with respect to the connection gives the torsional condition
\beq
{T^I}_{JK}=-3i\,k\alpha \,\bar{\psi}\gamma_5\gamma_M \psi\,{\epsilon^{MI}}_{JK}\,.
\eeq
The variation with respect to the spinor fields give the ordinary Dirac equation when the above is taken into account, and the variation with respect to the tetrad gives the Einstein-equations with an effective 4-fermion interaction coming from the torsional term in the field strength. 

We now wish to construct the symplectic form. Since the boundary term contains no derivatives, it does not contribute to the symplectic form. Taking the boundary of the manifold to be $\Sigma_1 \cup \Sigma_2\cup \tau_{\infty}$ where $\Sigma_1$ and $\Sigma_2$ are initial and final spacelike hypersurfaces and $\tau_{\infty}$ is the timelike cylinder at infinity, an arbitrary variation of the action on the covariant phase space reduces to:
\beq
\dl S=\bm{J}_{\partial M}+\bm{\theta}-\frac{1}{k}\int_{\tau_{\infty}}\star \dl (e\,e)\,\omega\,\label{Variation}
\eeq
where
\beq
\bm{J}_{\partial M}=-\frac{1}{k}\int_{\partial M}\star \omega\,\dl e\,e +\frac{\alpha}{2}\int_{\partial M}\dl \bar{\psi} \star e\,e\,e\, \psi -\bar{\psi} \star e\,e\,e\,\dl \psi.
\eeq
and $\bm{\theta}=\bm{\theta_\omega}+\bm{\theta_e}+\bm{\theta_\psi}+\bm{\theta_{\bar{\psi}}}$ with
\beqa
\bm{\theta_e}&=&\frac{1}{k}\int_M \dl e \left(e\star R-\star R\,e\right)\nn\\
& & +\alpha \int_M \frac{1}{2}\left(\bar{\psi}\star \dl (e\,e\,e) D\psi+ D\bar{\psi} \star \dl (e\,e\,e)\psi\right) -m\,\bar{\psi}\psi\star \dl(e\,e\,e\, e) \nn\\
\bm{\theta_\omega}&=&\frac{1}{k}\int_M \dl\omega\,D(\star\, e\,e)+\frac{\alpha}{2}\int_M \bar{\psi}\left(\star e\,e\,e\,\dl\omega -\dl\omega \star e\,e\,e \right)\psi \nn\\
\bm{\theta_\psi}&=&\alpha \int_M \left(\frac{1}{2}\left( D(\bar{\psi}\star e\,e\,e)+(D\bar{\psi})\star e\,e\,e \right)-m\bar{\psi}\star e\,e\,e\,e \right)\dl \psi \nn\\
\bm{\theta_{\bar{\psi}}}&=& \alpha\int_M \dl\bar{\psi}\left(\frac{1}{2}\left(\star e\,e\,e\,D\psi - D(\star e\,e\,e \,\psi)\right)-m \star e\,e\,e\,e \,\psi \right)
\eeqa
As usual, for arbitrary and independent variations, $(\dl e,\dl\omega,\dl\psi,\dl\bar{\psi})$, the vanishing of the individual terms $\bm{\theta_{\{i\}}}$ yield the pointwise equations of motion. Treating $\dl$ as an exterior derivative on the phase space, the exterior derivative of the symplectic one-form $\bm{J}$ yields the (pre)symplectic form:
\beq
\bm{\Omega}_{\partial M} \equiv -\dl \bm{J}_{\partial M}= \frac{1}{k}\int_{\partial M}\star \dl \omega \wedge \dl (e\,e)+\frac{\alpha}{2}\int_{\partial M}\dl(\bar{\psi}\star e\,e\,e)\wedge \dl\psi+\dl{\bar{\psi}}\wedge \dl(\star \,e\,e\,e\,\psi)\,.
\eeq
The ``covariant phase-space'', $\widetilde{\Gamma}$, is the subspace, $\phi:\widetilde{\Gamma}\rightarrow \Gamma$ of the phase space whose points consist of solutions to the equations of motion. On $\widetilde{\Gamma}$, the one-form $\bm{\theta}$ is identically zero. With appropriate boundary conditions defining the phase space, the symplectic flux across the timelike surface at infinity can be shown to vanish \cite{AshtekarSloan}. Restricting variations to those that preserve the equations of motion forces the $\bm{\theta}$ term to vanish and a second variation of the action on the covariant phase space can then be shown to yield the conservation equation:
\beq
\dl (\phi^*\dl S)\stackrel{\widetilde{\Gamma}}{=}\bm{\Omega}_{\Sigma_2}-\bm{\Omega}_{\Sigma_1}=0\,.
\eeq
Now, consider the action of a locally infinitesimal $Spin(3,1)$ gauge transformation generated by $\lambda$. Under such a transformation, the dynamical fields transform according to
\beqa
e &\rightarrow & e'=e+[\lambda, e]\nn\\
\omega &\rightarrow &\omega'=\omega -D_{\omega} \lambda\nn \\
\psi &\rightarrow & \psi'=\psi +\lambda \psi \nn\\
\bar{\psi} &\rightarrow & \bar{\psi}'=\bar{\psi}-\bar{\psi}\lambda \,.
\eeqa
A generic functional of the dynamical variables will then transform as follows:
\beqa
f[e,\omega, \psi, \bar{\psi}] \rightarrow f[e',\omega', \psi', \bar{\psi}']=f+\bm{\mathcal{L}_{\bar{\lambda}}}f \,,
\eeqa
where $\bm{\bar{\lambda}}$ is the vector field in the tangent space of the covariant phase space given by:
\beq
\bm{\bar{\lambda}}f=\int_{\Sigma}[\lambda, e] {\ts \frac{\dl f}{\dl e}}-D_{\omega}\lambda {\ts \frac{\dl f}{\dl \omega}}+ {\ts \frac{\dl f}{\dl \psi}}\lambda \psi-\bar{\psi}\lambda {\ts \frac{\dl f}{\dl \bar{\psi}}}\,.
\eeq
The symplectic form will be invariant under this transformation only if $\bm{\mathcal{L}_{\bar{\lambda}}\Omega}=\dl \bm{\Omega(\bar{\lambda},\ )}=0$. This means that $\bm{\Omega(\bar{\lambda},\ )}$ must be closed or, locally, exact. The local functional serving as the generator of local Lorentz transformation is the Gauss functional:
\beq
G(\lambda)=\frac{1}{k}\int_\Sigma-D\lambda\,\star e\,e +\frac{\alpha}{2}\int_\Sigma -\bar{\psi}\{\lambda,\star\,e\,e\,e\}\psi\,.
\eeq
As to be expected, the functional $G(\lambda)$ and $\bm{\bar{\lambda}}$ are canonical duals in the sense that
\beq
\bm{\Omega(\bar{\lambda},\ )}=\dl G(\lambda)\,,
\eeq
demonstrating that the symplectic form is invariant under local $Spin(3,1)$ transformations. 

Unlike the case of a closed spatial manifold $\Sigma$, the Gauss functional is no longer constrained to vanish --- taking the boundary to be the two-sphere at spatial infinity, $\partial \Sigma=S^2$, when evaluated on the the solutions to the equations of motion, the Gauss functional yields the boundary term (which was also noticed in the context of BF theory in \cite{Major:Surface} and asymptotically AdS spaces in \cite{Zanelli:AdSCharges}):
\beq
G(\lambda)\stackrel{\widetilde{\Gamma}}{=}-\frac{1}{k}\int_{S^2} \star\lambda\,e\,e =-\frac{1}{4k}\int_{S^2}\epsilon_{IJKL}\lambda^{IJ}e^K\,e^L
\eeq
This is in direct contrast to specific claims in the literature regarding the second order formalism \cite{Teitelboim:ECSurface,Teitelboim:Fermions,Teitelboim:Supergravity}, where it was stated that due to the tensorial transformation properties of the tetrad, the Gauss functional does not pick up a boundary term and retains its status as a constraint. 

We would like to study the properties of the boundary term $G(\lambda)$ evaluated on the covariant phase space.
We first note that given two different generators $\lambda_1$ and $\lambda_2$, the commutator of the vector fields $\bm{\bar{\lambda}_1}=\bm{\bar{\lambda}}(\lambda_1)$ and $\bm{\bar{\lambda}_2}=\bm{\bar{\lambda}}(\lambda_2)$ is a reflection of the $\mathfrak{spin}(3,1)$ Lie-algebra:
\beq
\bm{[\bar{\lambda}_1,\bar{\lambda}_2]}=\bm{\bar{\lambda}}([\lambda_1,\lambda_2])\,.
\eeq
This algebra is reflected in the Poisson algebra of $G(\lambda)$:
\beqa
\{G(\lambda_1), G(\lambda_2)\}\equiv \bm{\Omega(\bar{\lambda}_1, \bar{\lambda}_2)}=G([\lambda_1, \lambda_2])
\eeqa
Thus, the functionals $\sigma(\lambda)$ satisfies the same commutation relations as the generators of relativistic angular momentum. As we will see, this functional must be added to the ordinary ADM expression for angular momentum in order to account for the {\it internal} spin of the asymptotically flat spacetime.

\section{Isometries and Rigid Gauge Transformations \label{sec:Noether}}
The total angular momentum of an asymptotically flat spacetime is intimately connected to its behavior under a global Lorentz boost or rotation. Since generically in the bulk, Einstein gravity has both diffeomorphism and local Lorentz symmetry as separate, unrelated local symmetries, we need to delineate a local gauge transformation from a global isometry, or \textit{rigid gauge symmetry}.

Let's first begin with a review of the global symmetries of the Dirac equation in a flat, torsion free spacetime, focusing on the particular aspects that will generalize to arbitrary asymptotically flat Einstein-Cartan spacetimes. In the Minkowski case, the Dirac equation is given by:
\beq
\gamma^I\,e^\mu_I\,D_\mu\psi=m\,\psi\,.
\eeq
We will work in the standard Cartesian basis where the tetrad is trivial, $e^\mu_I=\delta^\mu_I$, and the spin connection coefficients are zero, $D_\mu=\partial_\mu$. Consider a diffeomorphism generated by a Killing vector $K^\mu={s^{\mu}}_\nu x^\nu$, where $s^{\mu\nu}=-s^{\nu\mu}$. To satisfy Killing's equation in this gauge, we need $\partial_\alpha s_{\mu\nu}=0$. Suppose that the spinor transforms actively by a simultaneous local Lorentz transformation and diffeomorphism given by $\psi\rightarrow \psi'=\psi+K^\mu \partial_\mu\psi + \lambda_{IJ}\frac{1}{4}\gamma^I \gamma^J \psi$, where $\lambda_{IJ}=-\lambda_{JI}$ is the infinitesimal boost or rotation parameter, which we will also assume to be constant in this gauge. The new spinor field $\psi'$ will also satisfy the same Dirac equation provided that the Killing vector and the local Lorentz transformation are related by:
\beq
\delta^\mu_I\, \delta^\nu_J\, \lambda^{IJ}\equiv \lambda^{\mu\nu}=-s^{\mu\nu}\,.
\eeq
This is an example of what we will refer to as a rigid gauge transformation. The key point is that the diffeomorphism an gauge transformation are related in such a way that the combination of the two gives a global isometry. Indeed, the above condition can be written in a more revealing way as follows. The desired transformation is nothing more than an ordinary combination of a diffeomorphism together with a local Lorentz transformation (an infinitesimal element of $Spin(3,1)\rtimes Diff_4$) such that $\psi\rightarrow \psi'=\psi+\mathcal{L}_{\bar{K}}\psi +\lambda \psi$ and $e^\mu_I \rightarrow e'^{\mu}_I =e^\mu_I +\mathcal{L}_{\bar{K}} e^\mu_I -{\lambda^K}_I e^\mu_K$. However, we impose the additional requirement that the transformation is a tetrad isometry in the sense that $e^\mu_I=e'^\mu_I$. Thus, the rigid gauge transformation must satisfy
\beq
\mathcal{L}_{\bar{K}} e^\mu_I ={\lambda^K}_I e^\mu_K\,, \label{KillingPair}
\eeq
which implies the conditions $\lambda^{\mu\nu}=-s^{\mu\nu}$ if $K^\mu={s^{\mu}}_\nu x^\nu$.

\subsection{The Noether Current of a Rigid Gauge Transformation}
Consider the flat-space (torsion-free) Dirac action in Cartesian coordinates
\beq
S=6 \alpha \int d^4 x \,\ts{\frac{1}{2}}\left(\bar{\psi}\gamma^\mu \partial_\mu \psi - \partial_\mu \bar{\psi} \gamma^\mu \psi\right) -m \bar{\psi}\psi
\eeq
Under an arbitrary variation, the action transforms on-shell by
\beq
\dl S \approx 6 \alpha \int d^4x\,\partial_\mu \left( \ts{\frac{1}{2}}\bar{\psi}\,\gamma^\mu \,\dl \psi - \ts{\frac{1}{2}}\dl \bar{\psi} \,\gamma^\mu \,\psi\right)
\eeq

Suppose now that the variation is a rigid gauge transformation 
\beqa
\dl \psi = \mathcal{L}_{\bar{K}}\psi +\lambda\psi \,. 
\eeqa
where $\bar{K}$ and $\lambda$ is a pair satisfying (\ref{KillingPair})
Under this transformation, the action is preserved on-shell:
\beqa
\dl S &=& 6 \alpha \int d^4 x\,\partial_\mu \left( K^\mu \left(\ts{\frac{1}{2}}\left(\bar{\psi}\gamma^\alpha \partial_\alpha \psi - \partial_\alpha \bar{\psi} \gamma^\alpha \psi\right) -m \bar{\psi}\psi \right)\right) \nn\\
&\approx & 0\,.
\eeqa
Putting the two expressions together, we have the conserved current
\beqa
j^{\mu}_{(\bar{K},\lambda)}=3 \alpha \left(K^\nu \left( \bar{\psi}\gamma^\mu \partial_\nu\psi -\partial_\nu \bar{\psi}\gamma^\mu \psi \right)
+\lambda_{\alpha\beta}\left(\bar{\psi} \{\gamma^\mu, \ts{\frac{1}{4}}\gamma^\alpha \gamma^\beta \}\psi\right)\right)\label{Noether}
\eeqa
which, in turn, yields the conserved charge
\beq
Q_{(\bar{K},\lambda)}=\int d^3 x \, j^0_{(\bar{K},\lambda)}\,.
\eeq 
To gain a better understanding for the conserved charges, we introduce a basis for the set of isometries of Minkowski space. It is convenient to introduce a new set of indices $\h{A}:=(\h{0},\h{1},\h{2},\h{3},\h{4})$ labelling the vector representation of the Lie algebra $\mathfrak{iso}(3,1)$ obtained from a contraction of $\mathfrak{so}(4,1)$. The set of ten basis vectors $\bar{K}^{\{\h{A}\h{B}\}}$ satisfy the following algebra under the Lie bracket:
\beq
\left[\bar{K}^{\{\h{A}\h{B}\}}\,,\,\bar{K}^{\{\h{C}\h{D}\}}\right]={f^{\{\h{A}\h{B}\}\,\{\h{C}\h{D}\}}}_{\{\h{E}\h{F}\}}\, \bar{K}^{\{\h{E}\h{F}\}}
\eeq
where ${f^{\{\h{A}\h{B}\}\,\{\h{C}\h{D}\}}}_{\{\h{E}\h{F}\}}$ are the structure constants for the Lie algebra $\mathfrak{iso}(3,1)$. For convenience, denoting the Lorentz indices by uppercase hatted indices towards the end of the alphabet, $\h{I},\h{J},\h{K}...=(\h{0},\h{1},\h{2},\h{3})$, a convenient basis of pairs $(\bar{K}^{\{\h{A}\h{B}\}}, \lambda^{\{\h{A}\h{B}\}})$ satisfying (\ref{KillingPair}) is:
\beqa
\bar{K}^{\{\h{I}\h{J}\}}= \lambda^{\{\h{I}\h{J}\}}_{IJ}\, {}^o e^I_\mu {}^oe^J_\nu\,x^\mu \frac{\partial}{\partial x_\nu} & \quad & \lambda^{\{\h{I}\h{J}\}}_{IJ}= \delta^{\h{I}}_I \delta^{\h{J}}_J-\delta^{\h{I}}_J \delta^{\h{J}}_I \label{generators1}\\
\bar{K}^{\h{I}}=\bar{K}^{\{\h{4}\h{I}\}}=\delta^{\h{I}}_\mu \frac{\partial}{\partial x_\mu} &\quad&  \lambda^{\{\h{4}\h{I}\}}_{IJ}= 0\,. \label{generators2}
\eeqa
With these definitions, we have a total of 10 conserved currents $j^\mu_{\{\h{A}\h{B}\}}$ yielding the ten conserved charges that we now identify with the energy-momentum and the (total=spin+orbital) angular momentum: 
\beqa
P^{\h{I}}&\equiv&\int d^3 x \,j^{0\,{\{\h{4}\h{I}\}}}=6 \alpha \int d^3 x \,\frac{1}{2}\delta^{\h{I}}_\mu \left(\bar{\psi}\gamma^0 \partial^\mu \psi -(\partial^\mu \bar{\psi})\,\gamma^0 \psi\right) \label{Noether1a}\\
J^{\h{I}\h{J}}_{tot} &=& L^{\h{I}\h{J}} +S^{\h{I}\h{J}} \equiv \int d^3 x \,j^{0\,\{\h{I}\h{J}\}}\nn\\&=&6 \alpha \int d^3 x\,  \frac{1}{2}\delta^{\h{I}\h{J}}_{\mu\nu}\left(\bar{\psi}\gamma^0 \,x^\mu\partial^\nu\psi -(x^\mu \partial^\nu\bar{\psi})\,\gamma^0 \psi+\bar{\psi}\{\gamma^0\,,\,\gamma^\mu\gamma^\nu\}\psi   \right)\,. \label{Noether1b}
\eeqa

A key property of the symmetry group can be seen from the form of the generators given above. The Killing vectors and the matrix generators $\lambda^{\{\h{I}\h{J}\}}$ generate metric isometries and internal Lorentz transformations. Together, they generate the group structure $Spin(3,1)\otimes (SO(3,1)\ltimes \mathbb{R}^{3,1})$ as will be discussed in more detail later. In addition to this, the generators of the internal boosts and rotations are not independent from the generators of the spacetime boosts and rotations, but instead are ``locked" by the condition (\ref{generators1}). Since the generators were constructed from the condition that the transformations preserve ${}^0e$, we will refer to the (component connected to the identity of) this group as the spin-enlarged Poincar\'{e} group and denote it $\mathfrak{G}({}^0e)$. This construct will be pivotal when considering the isometries of asymptotically flat spacetimes in later sections.

\subsection{Rigid Gauge Transformations with Curvature and Torsion}
We now wish to generalize the concept of a rigid gauge transformation to spacetimes with curvature and torsion, in a coordinate independent manner. For this it will be easiest to consider the transformation of the co-frame $e^I=e^I_\mu\,dx^\mu$, and the spin connection $\omega^{IJ}={\omega^{IJ}}_{\mu}\,dx^\mu$. The complete geometry is fixed in a local trivialization by specification of the co-frame and the spin connection. Therefore we wish to consider an infinitesimal transformation that does not affect these fields. Under a generic infinitesimal $Spin(3,1)\rtimes Diff_4$ transformation we have:
\beqa
e^I &\rightarrow & e'^I=e^I+\mathcal{L}_{\bar{K}}e^I+{\lambda^{I}}_K\,e^K \nn\\
\omega^{IJ}&\rightarrow & \omega'^{IJ}=\omega^{IJ}+\mathcal{L}_{\bar{K}}\omega^{IJ}-D_{\omega}\lambda^{IJ}\,.
\eeqa
A rigid gauge transformation, or a geometric isometry is defined by the conditions $e^I=e'^I$, and $\omega^{IJ}=\omega'^{IJ}$ yielding the defining conditions
\beqa
\mathcal{L}_{\bar{K}}e^I &=& -{\lambda^{I}}_K\,e^K \label{Rigid1a}\\ 
\mathcal{L}_{\bar{K}}\omega^{IJ} &=& D_{\omega}\lambda^{IJ} \label{Rigid1b} \,.
\eeqa
In the case of flat curvature and zero torsion, these conditions reduce down to the global isometries of the Dirac equation presented above. From the definition of the metric tensor, $\mathfrak{g}=\eta_{IJ} e^I \otimes e^J$, condition (\ref{Rigid1a}) requires that 
\beq
\mathcal{L}_{\bar{K}}\mathfrak{g}=0 \,,
\eeq 
so $\bar{K}$ must be a Killing vector. Thus, the isometry conditions are highly restrictive, and not all spacetimes will admit fields $\bar{K}$ and $\lambda$ such that (\ref{Rigid1a}) and (\ref{Rigid1b}) are satisfied. To analyze the following, it will be useful to introduce the concept of a gauge covariant Lie derivative. The generalization is straight-forward. We define the gauge covariant Lie derivative, $\widetilde{\mathcal{L}}_{\bar{K}}$, of a Lorentz vector (or spinor) valued p-form $\alpha$ such that under the transformation $\alpha\rightarrow g( \alpha) $, we have $\GLie{K}(g(\alpha))=g (\GLie{K} \alpha)$, and $\GLie{K} D_\omega\alpha=D_\omega\GLie{K} \alpha +\GLie{K}\omega\, \alpha -(-1)^{p}\alpha\,\GLie{K}\omega$. It can easily be shown that these two conditions are satisfied by
\beqa
\GLie{K}\alpha &=& (D \alpha)(\bar{K}) +D(\alpha(\bar{K})) \nn\\
\GLie{K}\omega &=& R(\bar{K})\,.
\eeqa
With these definitions, the conditions (\ref{Rigid1a}) and (\ref{Rigid1b}) can be written in a more manifestly covariant way by:
\beqa
\GLie{K}e^I &=&T^I(\bar{K})+DK^I= -{\ulambda^{I}}_K\,e^K \label{Rigid2a}\nn\\ 
\GLie{K}\omega^{IJ} &=& R(\bar{K})=D\ulambda^{IJ} \label{Rigid2b}
\eeqa
where $\ulambda^{IJ}=\lambda^{IJ}-\omega^{IJ}(\bar{K})$. Since the above expressions are manifestly covariant, it is sometimes convenient to work with the covariant Lie derivative. However, one must be careful in that fixing $\lambda$ to be phase space independent forces $\ulambda$ to be phase space dependent\footnote{From generic arguments, it can be argued that $\lambda$ and not $\underline{\lambda}$ should be phase space independent. For example, fixing $\dl\bar{K}=0$ and $\dl{\lambda}=0$, means that the variations $\dl e$ and $\dl\omega$ will themselves satisfy the rigid gauge conditions (covariant or non-covariant). This is not true if one tried to fix $\dl\bar{K}=0$ and $\dl{\underline{\lambda}}=0$}. The advantage to this covariant form is that in the flat space torsion-free limit, the generators $\widetilde{\mathcal{L}}_{\bar{K}^{\{\h{I}\h{J}\}}}$ generate the $SO(3,1)\ltimes \mathbb{R}^{3,1}$ subgroup and $\underline{\lambda}^{\{\h{K}\h{L}\}}$ satisfy the $Spin(3,1)$ algebra separately. The direct product between the two groups in $\mathfrak{G}({}^oe)$ is then manifest since (\ref{Rigid2b}) implies $\widetilde{\mathcal{L}}_{\bar{K}^{\{\h{I}\h{J}\}}}\underline{\lambda}^{\{\h{K}\h{L}\}}=0$. However, since the generators $\underline{\lambda}$ are phase space dependent, there are subtleties in the canonical theory, which roughly speaking come from the difficulty of generically separating spin angular momentum from orbital angular momentum in an arbitrary gauge\footnote{The difficulty stems from the the fact that generically the vector field $\bm{\underline{\bar{\lambda}}}=\frac{1}{k}\int -D\ulambda \,\frac{\dl}{\dl\omega}+[\ulambda,e]\frac{\dl}{\dl e}$ is generically not a symplectomorphism since $\bm{\Omega(\underline{\bar{\lambda}})}=-\dl \frac{1}{k}\int \star \ulambda\,e\,e +\frac{1}{k}\int \star\dl\ulambda\,e\,e$. Combined with a generator for the covariant Lie derivative, under the condition of a rigid gauge transformation, the extra terms cancel, so the combination is a symplectomorphism. However, it is therefore clear that it is difficult to generically separate the spin from the orbital angular momentum. These problems disappear in an asymptotically Cartesian gauge.}. To get around these subtleties, we will always work in a Cartesian basis at asymptotic infinity where $^{0}e^I_\mu$ is a constant so that $\p_\mu{}^{0}e^I_\nu=0$. In this case, the six boost and rotation Killing vectors are always paired with the six independent internal generators $\lambda$, and the corresponding generators to the translational Killing vectors are all zero. Furthermore, in this gauge the covariant and ordinary Lie derivatives coincide and one can easily separate spin from orbital angular momentum in the canonical theory.

Now, suppose we have a solution to the full set of the Einstein-Cartan-Dirac equations characterized by the fields $e$, $\omega$, $\psi$ and $\bar{\psi}$. Assume that the geometry admits a passive isometry of the form (\ref{Rigid2a}) and (\ref{Rigid2b}). Then given the solution $\psi$, the actively transformed field $\psi'=\psi+\Lie{K}\psi+\lambda \psi$ is also a solution \textit{for the same gravitational field configuration $e$ and $\omega$}.

With the definition of a rigid gauge transformation given above, the Noether charge can now be easily generalized to spacetimes with curvature and torsion. Suppose the spacetime $(M, e)$ admits a rigid isometry generated by the pair $(\bar{K},\lambda)$ satisfying (\ref{Rigid1a}) and (\ref{Rigid1b}). Then the Noether charge (evaluated on-shell) associated with this rigid isometry is 
\beq
Q_{\{\bar{K},\lambda\}}=\frac{\alpha}{2}\int_{\Sigma}\Lie{K}\bar{\psi}\star e\,e\,e \,\psi -\bar{\psi}\star e\,e\,e\, \Lie{K}\psi -\bar{\psi}\{\lambda,\star e\,e\,e \}\psi\,. \label{Noether2}
\eeq
In the low curvature, low spin density limit when the spacetime is approximately flat and torsion free, the spacetime admits ten Noether charges and (\ref{Noether2}) reduces to (\ref{Noether1a}) and (\ref{Noether1b}) for the ten basis generators.

\section{Relation between the Noether charge, the asymptotic ADM integral and the Komar integral \label{sec:Komar}}

In this section we will relate the Noether current corresponding to total angular momentum to a geometric integral on the boundary, \`{a} la Komar (see e.g.\cite{Carroll,Wald}). The construction of the integral identity will also clarify the relation between the Komar integral for the total angular momentum to the asymptotic ADM-like integral expression will we construct in proceeding sections. As usual in the Komar integral identities, we must assume the existence of a global (rigid) isometry to relate a matter integral in the bulk to a geometric integral on the boundary. The key ingredient allowing for such an integral identity for the total angular momentum of a Dirac field is the addition of an internal spin component to the geometric boundary integral representing the total angular momentum. 

To this end, consider a vector field $\bm{\bar{W}}$ on the tangent space $T\Gamma$. We will generically denote the components of the vector field by $\bm{\bar{W}}=(\delta e,\delta \omega, \delta\psi, \delta\bar{\psi})$. The vector field corresponding to a combination of an infinitesimal $Spin(3,1)\rtimes Diff_4$ transformation is given by
\beq
\bm{\bar{W}}=(\Lie{K}e+[\lambda, e]\,,\, \Lie{K}\omega -D_{\omega}\lambda \,,\, \Lie{K}\psi +\lambda \psi\,,\, \Lie{K}\bar{\psi}-\bar{\psi}\lambda)\,.
\eeq

Now consider the expression 
\beq
\bm{\Omega(\bar{W},\ )}=?? \label{KomarSymplectic}
\eeq
where we have yet to evaluate the right hand side. Since the vector $\bm{\bar{W}}$ is a bulk symmetry of the action, it also is the generator of a symplectomorphism so that the above expression is a closed (locally exact) one-form. Let's now evaluate the above expression using two different inputs for the field configuration. First we will evaluate the integral on the covariant phase space $\widetilde{\Gamma}\subset \Gamma$ consisting only of solutions to the Einstein-Cartan-Dirac field equations. We will also assume that we are restricting the phase space to the set of asymptotically flat spacetimes, the precise definition of which we will make clear later. For now, it will suffice to say that in the asymptotic limit, the frame and spin connection tend toward a fixed fiducial frame and spin-connection, ${}^0e$ and ${}^0\omega$, with ${}^0R=0$ and ${}^0T=0$. On this subspace, only variations are allowed that preserve the field equations. To derive the desired integral identities, we must fix a global vector field $\bar{K}$ and a global $\mathfrak{spin}(3,1)$ generator, $\lambda $. These are fixed in the sense that $\dl\bar{K}=\dl\lambda=0$. Evaluating (\ref{KomarSymplectic}) on this subspace, we have
\beq
\bm{\Omega(\bar{W},\ )}\stackrel{\widetilde{\Gamma}}{=} \frac{1}{k}\int_{\p \Sigma} \star \dl\omega \,[K,e]+\star \omega(\bar{K})\,\dl \,e\,e-\dl\left(\frac{1}{k}\int_{\p \Sigma}\star \lambda \,e\,e \right)\,.
\eeq
where $K=e(\bar{K})$. It can be shown that the first two terms on the right--hand side form a closed, locally exact one form allowing us to pull the $\dl$ outside of the integral.

The next step is to evaluate the expression on the set of field configurations that admit a rigid isometry generated by the vector field $\bar{K}$, which is now a Killing field, and its canonical pair $\lambda$ chosen so that $\Lie{K}e=-[\lambda ,e]$ and $\Lie{K}\omega=D\lambda $ for all tetrads and spin connections in this subspace, denoted $\Gamma_{\lambda}^{\bar{K}}$. The vector field, $\bm{\bar{W}}$ now reduces to 
\beq
\bm{\bar{W}}\stackrel{\Gamma_{\ulambda}^{\bar{K}}}{=}(0\,,\, 0\,,\, \Lie{K}\psi +\lambda \psi\,,\, \Lie{K}\bar{\psi}-\bar{\psi}\lambda)\,,
\eeq
so the desired expression depends only on the spinor-matter component of the symplectic form. Evaluated on this subspace, we have
\beqa
\bm{\Omega(\bar{W},\ )}&\stackrel{\Gamma_{\lambda}^{\bar{K}}}{=}& \dl\left(\frac{\alpha}{2}
\int_{\Sigma}\Lie{K}\bar{\psi}\star e\,e\,e\,\psi-\bar{\psi}\star e\,e\,e\, \Lie{K}\psi -\bar{\psi} \{\lambda\,,\,\star e\,e\,e\}\psi\right)\nn\\
& & +\frac{\alpha}{2}\int_{\Sigma} \Lie{K}\left(\bar{\psi}\star e\,e\,e \,\dl\psi -\dl\bar{\psi}\star e\,e\,e\,\psi \right)\,.\label{KomarIntermediate}
\eeqa
In evaluating the above expression we have assumed that the spinor fields fall off sufficiently rapidly outside a compact region that we can ignore boundary terms involving the spinor fields.

We can now combine these two expressions by considering the intersection $\widetilde{\Gamma}_{\lambda}^{\bar{K}}=\widetilde{\Gamma}\bigcap \Gamma_{\lambda}^{\bar{K}}$, consisting if the set of solutions to the Einstein-Cartan-Dirac equations that admit a rigid isometry generated by $\bar{K}$ and $\lambda$. In deriving the following, it will be useful to first point out the on-shell identities
\beqa
&\frac{\alpha}{2}\int_{\Sigma} \Lie{K}\left(\bar{\psi}\star e\,e\,e \,\dl\psi -\dl\bar{\psi}\star e\,e\,e\,\psi \right) 
 \stackrel{\widetilde{\Gamma}_{\lambda}^{\bar{K}}}{=}  0 & \label{IntegralIdentity2}\nn\\
&\frac{1}{k}\int_{\partial \Sigma}\iota_{\bar{K}}\,(\star\,\dl\omega\,e\,e)  \stackrel{\widetilde{\Gamma}_{\lambda}^{\bar{K}}}{=}  -\dl \left(\frac{\alpha}{2}\int_{\Sigma} \iota_{\bar{K}}(m\,\bar{\psi}\star\,e\,e\,e\,e\, \psi)\right)& \,. \label{IntegralIdentity3}
\eeqa
The first of these holds for any vector field $\bar{K}$ (not just Killing vectors) and establishes that not only is the total symplectic form hypersurface invariant, but the gravitational and fermionic components are separately invariant. Using these identities, on $\widetilde{\Gamma}_{\lambda}^{\bar{K}}$ we have
\beqa
\frac{1}{k}\int_{\p \Sigma} \star \dl\omega \,[K,e]+\star \omega(\bar{K})\,\dl \,e\,e-\dl\left(\frac{1}{k}\int_{\p \Sigma}\star \lambda \,e\,e\right)\stackrel{\widetilde{\Gamma}_{\lambda}^{\bar{K}}}{=} \dl Q_{\{\bar{K},\lambda\}} \label{IntegralIdentity1}
\eeqa
where $Q_{\bar{K},\lambda}$ is the Noether charge (\ref{Noether2}). Following \cite{AshtekarSloan2}, when the phase space is defined appropriately, the $\dl$ can be pulled out of the integral at asymptotic infinity to yield the following\footnote{Technically the equality is only valid up to a functional constant, but one can make general arguments that this constant should be set to zero, as we have done here.}
\beq
-\frac{1}{k}\int_{\p\Sigma}\star\,[K,e]\,\omega -\frac{1}{k}\int_{\partial \Sigma}\star\,\lambda\,e\,e = Q_{\{\bar{K},\lambda\}}\,.
\eeq
In the case that vector field is an asymptotic translational Killing vector $\bar{K}=\bar{T}$, the generators $\lambda$ vanish (in a gauge where ${}^0e^I_{\mu}$ is constant) at asymptotic infinity and the integral reduces to
\beq
P(\bar{T})\equiv -\frac{1}{k}\int_{\p\Sigma}\star\,[T,e]\,\omega = Q_{\{\bar{T},\lambda\}}\,. \label{Momentum}
\eeq
For an asymptotic rotational or boost Killing vector, $\bar{K}=\bar{L}$, the generators $\lambda$ do not vanish and the integral yields the relativistic momentum
\beq
L(\bar{L})+S(\lambda)\equiv -\frac{1}{k}\int_{\p\Sigma}\star\,[L,e]\,\omega -\frac{1}{k}\int_{\Sigma}\star\,\lambda\,e\,e =Q_{\{\bar{L},\lambda\}}\,. \label{AngMomentum}
\eeq
From this expression it is abundantly clear why we had to add the extra $\lambda$-dependent term to the expression for the total angular momentum -- in the expression for the Noether current, the term corresponding to the intrinsic spin angular momentum also involves $\lambda$ where it enters into the experession coupled to the intrinsic spin of the Dirac field. Thus, without this term the surface integral will not pick up the total angular momentum including intrinsic spin.

It is of interest to note using the Einstein-Cartan equation of motion expressing the torsion in terms of the axial current, we can rewrite the right hand side of the above to yield:
\beqa
-\frac{1}{k}\int_{\partial \Sigma}\star[K,e]\,\omega+\star \lambda\,e\,e  &\stackrel{\widetilde{\Gamma}_\lambda^{\bar{K}}}{=}& \frac{\alpha}{2}
\int_{\Sigma} \Lie{K}\bar{\psi}\star e\,e\,e\,\psi-\bar{\psi}\star e\,e\,e \,\Lie{K}\psi  \nn\\
& & -\frac{1}{k}\int_\Sigma \star\lambda(T\,e-e\,T) \label{Komar2}\,,
\eeqa
suggesting that the additional term in the surface integral of the left hand side comes directly from the non-zero torsion in the bulk that arises from the spin-torsion coupling. However, the precise role of that the torsion in the bulk plays in determining the geometry on the boundary is unclear at the present stage since the full results of this paper are strictly only known to be valid in the first order framework.

\subsection{The Komar integral} 
To establish the relation of the above integrals with the Komar integral identity, we need to evaluate (\ref{IntegralIdentity1}) in a slightly different manner. We first note that at asymptotic infinity where the torsion  vanishes we have $Tr(\star \ulambda\,e\,e)=\frac{1}{2}*d\widetilde{K}$ where $*$ is the external dual on the spacetime indices and $\widetilde{K}=K_{I}\,e^I$. Note that this expression involves $\ulambda=\lambda-\omega(\bar{K})$. Next we can rewrite the left hand side of (\ref{IntegralIdentity1}) to give 
\beq
-\dl \left(\frac{1}{k}\int_{\p\Sigma}\star\,\ulambda\,e\,e \right)-\frac{1}{k}\int_{\p\Sigma}\iota_{\bar{K}}\left(\star\dl\omega\,e\,e\right) \stackrel{\widetilde{\Gamma}_\lambda^{\bar{K}}}{=} \dl Q_{\{\bar{K},\lambda\}}\,.
\eeq
Using the identity (\ref{IntegralIdentity3}), we have the desired Komar integral identity:
\beq
-\frac{1}{2k}\int_{\p\Sigma}*d\widetilde{K} \stackrel{\widetilde{\Gamma}_\lambda^{\bar{K}}}{=} Q_{\{\bar{K},\lambda\}} -\frac{\alpha}{2}\int_{\Sigma}\iota_{\bar{K}}\left(m\,\bar{\psi}\star e\,e\,e\,e\,\psi \right)\,.
\eeq

To summarize, we have established the relation between the Noether current in the bulk, the ADM--like asymptotic expression for the momentum and angular momentum as a geometric boundary integral, and the Komar integral. Essential in the derivation of these expression is the addition of the term $\int_{\p\Sigma}\star\,\lambda\,e\,e $ to the boundary geometric integral. Without this term, we fail to pick up the internal spin contribution to the total angular momentum, and the identities no longer hold.

\section{Spacetimes with Asymptotic Killing Vectors \label{sec:Asymptotics}}
In deriving the integral identities above we had to begin with the assumption that the allowed spacetimes admit at least one global Killing vector $\bar{K}$ and its pair $\lambda$. A generic spacetime need not have any global Killing vectors. Since we are interested in asymptotically flat spacetimes, we require that the phase space consists of spacetimes with asymptotic Poincar\'{e} symmetry. More specifically, we fix a fiducial Lorentz frame ${^0e}$ and an associated connection $^0\omega$ such that ${}^0T=0$ and ${}^0R=0$.

Beig and Schmidt have shown that for space-times which are asymptotically flat the line element can be expressed as:
\be
\dd s^{2}=\left(1+\f{2\sigma} {\rho}\right)\, \dd\rho^{2} \,+\,\left(1-\f{2\sigma} {\rho}\right)\, \rho^2\, h_{ab} \,\dd\Phi^a \dd\Phi^b \,+\,o(\rho^{-1})\, , \label{Beig-Schmidt}
\ee
where $\sigma=\sigma(\Phi^I)$ is a function of the angular coordinates, $\rho$ the hyperbolic radius and $h_{ab}$ the metric on the unit hyperboloid. As was discussed in \cite{AshtekarSloan}, $\sigma$ is chosen to be reflection symmetric to eliminate super-translation ambiguities. Therefore our phase space is chosen to be the space of co-frames and connections compatible with this form. Let us therefore pick once and for-all a fiducial or `background' frame, ${}^0e$ compatible with the Minkowski metric described in the $\rho \rightarrow \infty$ limit above: ${}^0g_{\mu\nu} = \eta_{IJ}\, {}^0\!e_\mu^I\, {}^0\!e_\nu^J$ where $\eta_{IJ}$ is the Cartan-Killing metric on $SO(3,1)$. The asymptotic structure of such spacetimes was explored thoroughly in \cite{AshtekarHansen} in which a conformal factor, $R=\rho^{-1}$ is used to show that the geometry of spatial infinity is that of a hyperboloid. The restriction to even parity $\sigma$ corresponds to considering the leading contribution to the Weyl tensor to come entirely from the electric part.

We will require that $e$ admits an expansion of order 2, and hence we express
\be 
e= {}^0 e+\f{^1e}{\rho}+\f{^2e}{\rho^2}+o(\rho^{-2}) 
\ee
whence we adopt the notation $o(\rho^n)$ to indicate terms such that $\rho^n o(\rho^{-n}) \rightarrow 0$ as $\rho \rightarrow \infty$. The cartesian coordinates $x^\mu$ of $g^o_{\mu\nu}$ and the associated radial-hyperboloid coordinates $(\rho, \Phi^i)$ will be used in asymptotic expansions.

Let us now define the phase space that we will be working with throughout this paper. From general considerations, the phase space will consist of a dynamical tetrad with strong constraints on its asymptotic fall-off that define our notion of asymptotic flatness. The tetrad must be compatible with the Beig-Schmidt metric, together with parity conditions imposed in \cite{AshtekarSloan} that eliminate the logarithmic and super-translation ambiguities. Although \cite{AshtekarSloan} worked in a tetrad framework, the internal $SO(3,1)$ gauge freedom was fixed at the level of $^1 e$. Thus, we require a slight generalization of the phase space defined there. Generically, one can show that a tetrad compatible with the fixed form of the metric (\ref{Beig-Schmidt}) with the required parity conditions implies that the next to leading term in the expansion of the tetrad has the form
\beq
\label{e1} {}^{1}{e_{\mu}^{I}} = \sigma(\Phi)\, (2\rho^{I} \rho_{\mu} -{}^0{e_{\mu}^{I}}) +\beta^I{}_\mu 
\eeq
in which we adopt the convention of converting between internal and space-time indices with ${}^0e$ and hence set $\rho^I = {}^0 e^{\mu I} \rho_\mu$, and we raise and lower internal indices with $\eta$. In this expression, the new contribution, $\beta^I{}_\mu$ must be such that $\beta_{IJ}\equiv {\beta_{I\mu}} {}^{0}e^\mu{}_J=-\beta_{JI}$. This additional term still preserves the required form of the metric. On the other hand, we will demand that $^{1}e$ retains even parity under the mapping of antipodal points of the asymptotic two-sphere. This enforces the constraint that $\alpha_{IJ}$ must also be of even parity.

We use the compatibility between co-frame and connection to set $^0\omega={}^1\omega=0$, as our metric should asymptotically approach Minkowski space, leaving only ${}^2 \omega$ to be determined completely by $\sigma$, $\beta$ and ${}^0 e$.
We note that the symmetries of both $^1e$ and $^2\omega$ are fixed by the symmetries of $\sigma$, $\beta$ and ${}^0 e$. Since $\sigma$ and $\beta$ are of even parity so is $^1e$ and in turn this sets $^2\omega$ to be of odd parity, facts which will be crucial in establishing the finiteness of our system. As was stated in \cite{AshtekarSloan} this restriction is not excessive -- it places no restriction on the 4-momentum or Lorentz momentum of the system, but suffices to allow for a well defined action principle and finite conserved charges.

Let us now turn our attention to the spin contribution to angular momentum in the total Hamiltonian. Generically, the allowable gauge transformations are the transformations that preserve the phase space defined above. To lowest order, the allowed diffeomorphisms must preserve the fiducial flat metric ${}^0g_{\mu\nu}$, which in fact fixes the first {\it two} orders of the diffeomorphism generators. The boosts and rotations are generated by the Killing vector ${}^{-1}\bar{K}$, which are characterized by their asymptotic behavior, ${}^{-1}\bar{K}\approx \rho$, and their parity: the components ${}^{-1}K^\mu$ must have parity odd, making the generator $\mathcal{L}_{{}^{-1}\bar{K}}$ an even operator under parity inversion. The translations are generated by the four translational Killing vectors ${}^0\bar{K}$, which are also characterized by their asymptotic behavior ${}^0\bar{K}\approx \rho^0$, and their parity: the components ${}^0K^\mu$ have even parity, making   $\mathcal{L}_{{}^0\bar{K}}$ an odd operator. 

We expand our spin gauge transformation similarly. Generically, an internal gauge transformation is given by $g(x)\in \mathcal{G}=Spin(3,1)$. Denoting $\mathcal{G}_0$ as the component of $\mathcal{G}$ connected to the identity, and $\mathcal{G}^0_0$ as the identity connected component of $\mathcal{G}$ whose elements fall off at least as fast as $g \sim \frac{1}{\rho^3}$, the elements of this subgroup $\mathcal{G}^0_0$ are the pure internal {\it gauge} transformations of the theory. The remaining elements of $\mathcal{G}_0/\mathcal{G}^0_0$ can be expanded $g={}^0g+{}^1g/\rho+{}^2g/\rho^2$. An infinitesimal element of this subspace, $g\approx 1+\chi$, is given by the Lie algebra element
\beq
\chi = {}^0\chi + \f{{}^1\chi}{\rho} + \f{{}^2\chi}{\rho^2}\,.
\eeq
From the requirement that our defined phase space is preserved under the allowable set of transformations, we can now impose constraints on the asymptotic fall-off of the allowable internal gauge transformations. As demonstrated in \cite{AshtekarSloan}, the constraints on the form of the metric at asymptotic infinity (that $\sigma$ is parity even), constrains the allowable diffeomorphisms at asymptotic infinity by eliminating the logarithmic and super translation ambiguities. The remaining set of diffeomorphisms are generated by $\bar{K}={}^{-1}\bar{K} + {}^{0}\bar{K}+o(\rho^0)$, where ${}^{-1}\bar{K}$ is a boost or rotation Killing vector (which grows like $\rho$) and ${}^{0}\bar{K}$ is a translational Killing vector (which goes like ${}^0\bar{K}\sim \rho^0$). These elements both preserve the form of ${}^0e^I_\mu$, which is constrained to be constant. This constraint also enforces ${}^0\chi$ to be a constant, global gauge transformation as well. On the other hand, under a generic transformation, we have
\beq
{}^{1}e \rightarrow {}^1{e'}={}^1 e +\mathcal{L}_{ {}^0\bar{K}} {}^0 e +\mathcal{L}_{ ^{-1}\bar{K}} {}^1e +[ {}^0\chi , {}^{1} e ] +[ {}^1\chi , {}^0e ] \ .
\eeq 
Since $\mathcal{L}_{ {}^{0}\bar{K}} {}^{0} e =0$ and $\mathcal{L}_{ {}^{-1}\bar{K}} {}^{0}e =-[{}^{0}\lambda , {}^0 e] $, for some ${}^0\lambda$, and noting that the hyperbolic radial vector is, by definition, preserved by boost and rotations so that $ \mathcal{L}_{ {}^{-1}\bar{K}}\rho^\mu =0$, we have
\beq
{}^1 e'^I= \sigma' (\rho^I d\rho -^{0} e^I)+{\beta' {}^I{}_K} {}^0 e^K
\eeq
with 
\beqa
{}^0 e' &=& {}^0 e+ \mathcal{L}_{{}^{-1}\bar{K}}{}^0e+[{}^0\chi,{}^0e] \nn\\ 
\sigma' &=& \sigma+\mathcal{L}_{ ^{-1}\bar{K}}\sigma \nn\\
\beta'  &=& \beta+\mathcal{L}_{ ^{-1}\bar{K}}\beta + [ ^0\chi,\beta]+^{1}\chi
\eeqa

Recalling that $\mathcal{L}_{ {}^{-1}\bar{K}}$ is an even operator (it sends parity even functions to parity even functions, and parity odd to parity odd), and ${}^{0}\chi$ must be constant, it is clear that ${}^0 e'$ is still constant, and $\sigma'$ still has even parity. On the other hand since we have constrained $\beta$ to be of even parity, the phase space is preserved if and only if ${}^1\chi$ is parity even as well. 

Let us summarize our phase space and the constraints it puts on the allowable symmetries. An arbitrary infinitesimal diffeomorphism can be expanded $\phi_{\bar{K}}=1+\mathcal{L}_{{}^{-1}\bar{K}}+\mathcal{L}_{{}^0\bar{K}}+o(\rho^0)$, and an arbitrary gauge transformation in $\mathcal{G}_0$ can be expanded $g={}^0 g + {}^1 g/\rho +{}^2 g /\rho^2+o(\rho^{-2})$. The phase space is then constrained as follows
\begin{itemize}
\item{${}^0e$ is a fiducial flat tetrad with constant components, ${}^0e^I_\mu$ in a Cartesian coordinate. This implies
${}^0 g$ is a constant, global gauge transformation.}
\item{${}^1 e$ is constrained to be compatible with the Beig-Schimdt form of the metric, in addition to being {\it parity even}. Specifically its form is given by (\ref{e1}). This constrains the first two diffeomorphism generators, ${}^{-1}\bar{K}$ and ${}^0\bar{K}$, to be Killing vectors of the fiducial flat metric, and it eliminates the logarithmic and super translation ambiguities. In addition, this constrains ${}^{1}g$ to be {\it parity even}.\footnote{With this modification of the phase space, technically one must re-analyze the symplectic structure to ensure its finiteness as well as the vanishing of the symplectic flux across the timelike surface at asymptotic infinity, which is necessary for conservation. Although we will not do this here, we note that even if these properties fail for arbitrary even $\beta^I{}_J$, one is still free to retreat to the stronger phase space condition that $\beta^I{}_J$ is {\it constant}. In this case we are ensured finiteness of the conserved symplectic from the results of \cite{AshtekarSloan} since the modified phase space here simply differs from the phase space considered there by allowing global gauge transformations, which does not alter the action. This stronger constraint on the phase space does not affect any of the results to follow, including the form of the conserved charges. However, for generality we will consider the case where $\beta^I{}_J$ is simply parity even.}}
\item{${}^2 e$ is left unconstrained. This imposes no restrictions on higher order diffeomorphisms or gauge transformations.} 
\end{itemize}

\subsection{The Gauss functional}
Let us now evaluate the Gauss functional with imposed conditions on the phase space.
From our boundary conditions we can form an expansion:
\beq
G(\chi)= {}^0 G+{}^1G+{}^2G +o(\rho^{-2})
\eeq
wherein the terms are defined:
\beqa
{}^0 G &=& -\frac{1}{k}\int_{S^2} \star {^0\chi}\,{^0e}\,{^0e} \nn\\
{}^1 G &=& -\frac{1}{k}\int_{S^2} \rho^{-1}( 2\star {^0\chi}\,{^0e}\,{{}^1e}+\star {{}^1\chi}\,{{}^0e}\,{{}^0e})\nn\\
{}^2 G &=& -\frac{1}{k}\int_{S^2} \rho^{-2}(\star {{}^0\chi}\,{{}^1e}\,{{}^1e}+2\star {{}^1\chi}\,{{}^0e}\,{{}^1e}+\star {{}^2\chi}\,{^0e}\,{{}^0e}+2\star {{}^0\chi}\,{{}^0e}\,{{}^2e})  \,.
\eeqa

In the limit $\rho\rightarrow \infty$, only these first three terms (potentially) contribute, the rest are identically zero. At this point we notice that the first two orders in the expansion are potentially divergent, so we need to evaluate these under the assumption of asymptotic flatness to ensure finiteness of the expression. It is tempting to consider restricting our allowable transformations to be those which only yield a finite contribution to $G(\chi)$, however it is apparent that simply using power counting arguments would not suffice for this purpose - these would rule out the sector of rigid rotations, those described by $\chi = {}^0\chi$. However we shall see that the potentially divergent terms are in fact zero by use of parity arguments.

\subsection{Evaluating the potentially divergent terms}

We employ parity arguments repeatedly to show that the potentially divergent terms in our expansion of $G(\lambda)$ are in fact zero. Let us first examine the term ${}^0 G$: 
\beq 
{}^0 G= -\frac{1}{k}\int_{S^2} \star {^0\chi}\,{^0e}\,{^0e} 
\eeq  
Since $^0e$ is of even parity and $^0\lambda$ is of odd parity the integrand, $\star {^0\lambda}\,{^0e}\,{^0e}$ is itself of odd parity, being the product of two even terms and one odd term. Hence we are integrating an odd function over $S^2$ which will evaluate to zero.

To understand this in some more detail, in a Cartesian coordinate system we write ${}^0 e^I={}^0 e^I_\mu \,dx^\mu$ and recall that the coefficients are constant. In addition, we recall that ${}^0\chi$ must be constant as well. The integral then becomes:
\beqa 
{}^0G &=& - \frac{1}{4k}\int_{S^2} \epsilon_{IJKL}  {}^0\chi^{IJ} \,{}^0 e^K\,{}^0 e^L \nn\\
&=& \frac{1}{k}\int_{S^2} {}^0\chi^{IJ}n_I \hat{\rho}_J \, \widetilde{\sigma}^{(2)} 
\eeqa
where $n^I\equiv {}^0e^I_\mu n^\mu$ is the timelike normal to the hypersurface, $\hat{\rho}^I$ is the unit hyperbolic radial vector, and $\widetilde{\sigma}^{(2)}$ is the area two-form on $S^2$ induced by the flat tetrad ${}^0 e$. Since $n_I$ and ${}^0\chi^{IJ}$ both have even parity, but $\hat{\rho}_J$ has odd parity, the total integrand (with respect to the measure $\widetilde{\sigma}^{(2)}$) has odd parity, and therefore integrates to zero on the domain $S^2$. 
 
In fact, the parity arguments generalize to most of the terms in the expansion: one can show that the single term $^{0}G$, both of the terms in $^1 G$, and the first two terms of $^2 G$ can be put in the form
\beq
\int_{\Sigma}\rho^{-n} X_{IJ} \,n^I\rho^J \,\widetilde{\sigma}^{(2)}
\eeq
where $\widetilde{\sigma}^{(2)}$ is the metric volume form on $S^2$ induced by $^0 e$ and $X_{IJ}$ has {\it even} parity. Again, since $n^I$ has even parity and $\hat{\rho}^J$ has odd parity, the total integrand has odd parity, so terms of this form integrate to zero.

Thus, all of the potentially divergent terms are identically zero, so $G(\lambda)$ is explcitly finite with its sole contribution being:
\be 
G(\chi)={}^2 G=-\frac{1}{k}\int_{S^2} \hat{\rho}^{-2}( \star {^0\chi}\,{^0e}\,{^2e}+\star {^2\chi}\,{^0e}\,{^0e}) 
\ee
This establishes the finiteness of the theory with the chosen phase space. This construction follows the spirit of that used for the orbital contribution in \cite{AshtekarSloan} in which parity arguments were required to show finiteness. Strictly speaking, we only need to consider the space of rigid rotations in this work, however since we have placed no restrictions on ${}^3\omega$ it is pleasant to note that similarly no restrictions need be placed on ${}^2\chi$. However, we will now show that the spin-enlarged Poincar\'{e} algebra follows strictly from the rigid rotation sector.

\section{Generating functionals of the Poincar\'{e} algebra \label{sec:Generators}}
In the absence of internal degrees of freedom, as in geometrodynamics, the key defining property of asymptotic flatness is the existence of a Poincar\'{e} symmetry at asymptotic infinity. Whereas the bulk gauge symmetry has the generic structure of $Diff_4(M)$, the gauge symmetry must reduce to the set of symmetries that preserves the fiducial flat metric, ${}^0g_{\mu\nu}$, near asymptotic symmetry. Thus, near asymptotic infinity, the gauge symmetries reduce to true symmetries, and the algebraic structure of the generators of the gauge transformations must reduce to Lie algebra of the subgroup $Diff_4(\eta)=ISO(3,1)$. Without this requirement, one loses the notion of asymptotic flatness and the conservation of momentum and orbital angular momentum.

With internal spin-degrees of freedom, the set of gauge symmetries is enlarged. In particular the bulk gauge group, the diffeomorphism group, is enlarged to the bulk gauge group $Spin(3,1)\rtimes Diff_4(M)$. The true symmetry group at asymptotic infinity is, thus, also modified. To see precisely how it is modified, we need only to look at the algebra generated by the rigid gauge transformations in the particular basis given by (\ref{generators2}). Since the set of vectors $\bar{K}^{\{\hat{I}\hat{J}\}}$ are Killing vectors of a fiducial flat metric, which in this basis is ${}^0g_{\mu\nu}=\eta_{\mu\nu}$, their algebra under the Lie bracket is the subalgebra $\mathfrak{iso}(3,1)$. Similarly, the generators $\lambda^{\{\hat{I}\hat{J}\}}=\frac{1}{4}\gamma^K\gamma^L \,\lambda^{\{\hat{I}\hat{J}\}}_{KL}$ also generate a subalgebra, namely $\mathfrak{spin}(3,1)$. A key property that is obvious in this basis, but also must hold in any basis, is that the differential generators of the subalgebra $\mathfrak{iso}(3,1)$ commute with the matrix generators of the subalgebra $\mathfrak{spin}(3,1)$:
\beq
\Lie{K}[\lambda \,,\, f]-[\lambda\,,\,\Lie{K}f]=[\Lie{K}\lambda \,,\,f]=0\,.
\eeq
Because of this the true rigid symmetry group loses its status of a semi-direct product and reduces down to
\beq
Spin(3,1)\otimes ISO(3,1)=Spin(3,1)\otimes (SO(3,1)\ltimes \mathbb{R}^{3,1})\,. \label{GroupStructure}
\eeq
In addition to this, the subgroups $Spin(3,1)$ and $SO(3,1)$ are not independent, but are constrained by the condition that they define a rigid gauge transformation, in this case preserving the fiducial flat tetrad ${}^0 e^I_\mu$. Thus there must exist a map from the fundamental representation of $\mathfrak{spin}(3,1)$ to the differential operator representation of $\mathfrak{so}(3,1)$, denoted by $\phi:M(\mathfrak{spin}(3,1))\rightarrow D(\mathfrak{so}(3,1))$. At the level of the Lie algebra, the map must be bijective (it should be 2 to 1 at the group level). Specifically, in this basis the map is given by
\beq
\phi\left(\ts{\frac{1}{4}}\gamma^K\gamma^L\,\lambda^{\{\hat{I}\hat{J}\}}_{KL}\right)= \lambda^{\{\h{I}\h{J}\}}_{KL}\,{}^0 e_\mu^K {}^0 e_\nu^L \,x^\mu \frac{\partial}{\partial x_\nu}=\bar{K}^{\{\h{I}\h{J}\}}\,.
\eeq
In total, we will denote the identity connected component of the group with structure (\ref{GroupStructure}) and the above locking condition by $\mathfrak{G}({}^0e^I_\mu)$.

The key point, is that the generators of symmetries at asymptotic infinity should also preserve this algebraic structure. If not, our notion of asymptotic flatness fails, and we will lose any notion of conserved total angular momentum. We can now easily generalize the algebraic restrictions on the generators of rigid gauge transformations to generators of symmetries at asymptotic infinity. First, the requirement that the gauge group, $Spin(3,1)\rtimes Diff_4(M)$ reduces to the subgroup $Spin(3,1)\otimes (SO(3,1)\ltimes \mathbb{R}^{3,1})$ must be carried over to the functional generators of momentum and angular momentum. Splitting total angular momentum into an orbital and a spin contribution, $J^{\hat{I}\hat{J}}_{total}=L^{\hat{I}\hat{J}}+S^{\hat{I}\hat{J}}$, under the Poisson bracket, the Lie algebra of $L^{\hat{I}\hat{J}}$ and the ordinary momentum generators $P^{\hat{K}}$ must reproduce the subalgebra $\mathfrak{iso}(3,1)$. Similarly, the spin generators $S^{\hat{I}\hat{J}}$ must reproduce the algebra $\mathfrak{spin}(3,1)$. These two requirements are automatically guaranteed from general arguments. However, the third requirement, that the symmetry group reduces to a direct, rather than semi-direct product is not guaranteed unless the term in $S^{\hat{I}\hat{J}}$ involving $^2\lambda^{\{\hat{I}\hat{J}\}}$ vanishes. More specifically, since ${}^{2}\lambda$ is an unconstrained function at asymptotic infinity if we simply take $S(\lambda)=G(\lambda)$, we will generically have
\beq
\{L(\bar{K}), S(\lambda)\}=S(\Lie{K} ({}^2\lambda))\neq 0\,,
\eeq
and the algebra will fail to be that of a direct product.

Secondly, the requirement that that there exists a map between the matrix generators of the internal group $Spin(3,1)$ and the differential operator generators of the spacetime isometry group $SO(3,1)$ easily generalized to the requirement that there exists a bijective map from functionals to functionals, $\Phi:\mathcal{F}\rightarrow \mathcal{F}$, such that as functionals\footnote{We emphasize here that this is a map from functionals to functionals. This does not mean that the spin angular momentum and the orbital angular momentum are numerically identified.} $\Phi(S^{\hat{I}\hat{J}})=L^{\hat{I}\hat{J}}$. Here we appeal to the results of (\cite{AshtekarSloan}) where it was shown that orbital angular momentum functional is explicitly finite, and all but the leading order terms in the diffeomorphism generators vanish by parity arguments and the restriction on the phase space, which we have employed here. Once again, this implies that the map exists only for the ${}^0\lambda$ term in $S^{\hat{I}\hat{J}}$. Thus, we {\it define} the total angular momentum to be the combination 
\beq
J^{\hat{I}\hat{J}}\equiv L^{\hat{I}\hat{J}}(\bar{K})+S^{\hat{I}\hat{J}}({}^0\lambda)
\eeq
where $\bar{K}$ and ${}^0\lambda$ are canonical pairs forming a rigid gauge transformation. We can now check that the resulting algebra has the desired properties.

First we note that with the stated conditions, the map $\Phi$ exists and has the pleasing form in the Cartesian basis for the vector generators (\ref{generators2}):
\beq
\Phi(S^{\hat{I}\hat{J}}(\lambda^{\{\hat{I}\hat{J}\}}))=L^{\hat{I}\hat{J}}(\phi(\lambda^{\{\hat{I}\hat{J}\}}))=L^{\hat{I}\hat{J}}(\bar{K}^{\{\hat{I}\hat{J}\}})\,.
\eeq
Next we need to check that the algebra of the momentum, orbital angular momentum, and spin angular momentum forms a representation of the Lie algebra of $\mathfrak{G}({}^o e)=Spin(3,1)\otimes (SO(3,1)\ltimes \mathbb{R}^{3,1})$. The algebra can be computed explicitly resulting in the following
\beqa
\{ L^{\hat{I}\hat{J}}\,,\, L^{\hat{K}\hat{L}}\}&=& 
2\eta^{\h{I}[\h{K}}L^{\h{L}]\h{J}}- 2\eta^{\h{J}[\h{K}}L^{\h{L}]\h{I}}  \nn\\
\{ L^{\hat{I}\hat{J}}\,,\, P^{\hat{K}}\}&=& 2 \eta^{\h{K}[\h{J}}P^{I]}\nn\\
\{P^{\hat{I}}\,,\,P^{\hat{J}}\} &=& 0\,, \label{G1}
\eeqa
which defines the subalgebra of $ISO(3,1)=SO(3,1)\ltimes \mathbb{R}^{3,1}$, together with 
\beqa
\{ S^{\hat{I}\hat{J}}\,,\, S^{\hat{K}\hat{L}}\}&=& 2\eta^{\h{I}[\h{K}} S^{\h{L}]\h{J}}- 2\eta^{\h{J}[\h{K}} S^{\h{L}]\h{I}}\nn \\
\{ S^{\hat{I}\hat{J}}\,,\, P^{\hat{K}}\} &=& 0 \nn\\
\{ S^{\hat{I}\hat{J}}\,,\, L^{\hat{K}\hat{L}}\} &=& 0\,. \label{G2}
\eeqa
In total this is precisely the Lie algebra of $\mathfrak{G}({}^o e)=Spin(3,1)\otimes (SO(3,1)\ltimes \mathbb{R}^{3,1})$.

Since the generating functionals themselves form a representation of the appropriate algebra, we can now construct two invariants from the Casimirs. As usual, the first invariant is quadratic Casimir which yields the mass square:
\beq
C_2\equiv -M^2=P_{\hat{I}}P^{\hat{I}}\,.
\eeq
The second invariant is the quadratic Casimir which we identify with the total (spin+orbital) angular momentum invariant of the spacetime:
\beq
C_4/C_2 \equiv W_{\h{I}}W^{\h{I}}/P_{\hat{I}}P^{\hat{I}}=S(S+1) \quad \quad \quad W_{\h{I}}=\frac{1}{2}\epsilon_{\h{I}\h{J}\h{K}\h{L}}P^{\h{J}}(L^{\h{K}\h{L}}+S^{\h{K}\h{L}})\,.
\eeq
The invariance of these quantities follows from the commutation relations (\ref{G1}) and (\ref{G2}).

\subsection{A new Poincar\'{e} invariant charge}
In the previous section, we saw that the from the canonical pair, $\{\bar{K}^{\{\hat{I}\hat{J}\}},{}^0\lambda^{\{\hat{I}\hat{J}\}}\}$ forming a rigid gauge transformation, one can construct the generating functionals of the spin-enlarged Poincar\'{e} algebra. On the other hand, given {\it any} infinitesimal gauge transformation, $g\approx 1+{}^0\lambda+{}^1\lambda/\rho+{}^2\lambda/\rho^2$, satisfying our constraints on $\mathcal{G}_0$ one can define a conserved charge as follows. First find the canonical dual Killing vector, $\bar{K}$ to ${}^0\lambda$, and construct the functional
\beq
L(\bar{K})+G({}^0\lambda+{}^1\lambda/\rho+{}^2\lambda/\rho^2)=L(\bar{K})+S({}^0\lambda)+Q({}^2 \lambda)
\eeq
where we have defined a new charge 
\beq
Q({}^2 \lambda)\equiv -\frac{1}{k}\int_{S^2} \rho^{-2}\star{}^2\lambda\,{}^0e\,{}^0e\,.
\eeq
This charge is manifestly finite, and it is conserved. Physically, it is the generator of the internal gauge symmetries $g\sim o(\rho^{-1})$. Furthermore, as we will now show, the charge is invariant under the spin-enlarged Poincar\'{e} trasnformations. The charge is peculiar in the sense that it only depends on the asymptotic structure of the gauge transformation\footnote{It can also depend on the choice of fiducial flat tetrad ${}^0e$, bit this dependence can be eliminated by a suitable basis invariant method of defining $g$}. In this sense, the integral $Q({}^2\lambda)$ picks out Poincar\'{e} invariant topological information associated with the global structure of the gauge transformation. 

We will now show that $Q({}^2\lambda)$ is invariant under the spin-enlarge Poincar\'{e} group, or equivalently that linear momentum and the {\it total} angular momentum are invariant under ${}^2\lambda$ transformations.
To see this, consider the change in the orbital piece of the angular momentum under a $^{2}\lambda$ transformation:
\beqa
\{Q({}^2\lambda)\,,\,L(\bar{K})\} &=& \bm{\Lie{K}} Q({}^2\lambda)\nn\\
&=&-\frac{1}{k}\int_{S^2} \rho^{-2}\left(\star \iota_{\bar{K}} ([{}^2\lambda, e\,e])\,\omega - \star \iota_{\bar{K}}(e\,e)\,D\,{}^2\lambda\right) \nn\\
&=& +\frac{1}{k}\int_{S^2} \rho^{-2}\,\star\, \iota_{\bar{K}}({}^0e\,{}^0 e)\,d\,{}^2\lambda\,.
\eeqa
Integrating by parts and recalling that $d\,{}^0 e=0$, this reduces to
\beq
\frac{1}{k}\int_{S^2}\rho^{-2}\,\star \Lie{\bar{K}}(^{0}e\,^{0}e)\,{^2}\lambda=\frac{1}{k}\int_{S^2}\rho^{-2}\,\star [{}^0\lambda,{}^2\lambda]\,^{0}e\,^{0}e 
\eeq
where we have used the fact that $\Lie{K}{}^0 e=-[{}^0 \lambda, {}^0 e]$. Thus in total under this transformation we have
\beq
Q({}^2\lambda)\rightarrow Q({}^2\lambda) +\frac{1}{k}\int_{S^2}\rho^{-2}\,\star [{}^0\lambda, {}^2 \lambda]\,e\,e\,.
\eeq

We conclude that the charge $Q({}^2\lambda)$ is not by itself invariant under the set of allowed diffeomorphisms.
On the other hand, if we consider the total expression $J_{tot}=L(\bar{K})+S({}^0\lambda)$ and note that under gauge part of the transformation we have
\beqa
Q({}^2 \lambda)&\rightarrow & Q({}^2\lambda)+\bm{\Lie{{}^0\lambda}}Q({}^2\lambda)\nn\\
&=& Q({}^2\lambda)-\frac{1}{k}\int_{S^2} \rho^{-2}\,\star [{}^0\lambda , {}^2\lambda]\, e\,e\,.
\eeqa
So, we see that the separate contributions from the change in the orbital and spin pieces cancel yielding
\beq
\{Q({}^2\lambda)\,,\,J_{tot}\}=0\,.
\eeq
In addition one can show that the linear momentum also commutes with $Q({}^2\lambda)$:
\beq
\{Q({}^2\lambda)\,,\,P(\bar{K})\}=0\,.
\eeq
Thus, the charge $Q({}^2\lambda)$ is now invariant under the spin-enlarged Poincar\'{e} group. We can turn this argument around to say that the {\it total} angular momentum is invariant under gauge transformations $g\sim o(\rho^{-1})$. This is yet another reason why the internal spin functional must be added to the total angular momentum -- it is necessary in order to preserve gauge invariance of the total angular momentum. 

In total the algebra of the conserved quantities has the formal structure:
\beq
\begin{array}{llll}
\{L\,,\,L\}\sim L \quad& \{S\,,\,S\}\sim S & \quad \{L\,,\,Q\}\sim Q & \quad\{Q\,,\,Q\} \sim 0\\
\{L\,,\,P\}\sim P \quad&  \{S\,,\,L\}\sim 0 & \quad\{S\,,\,Q\}\sim -Q &  \\
\{P\,,\,P\}\sim 0 \quad& \{S\,,\,P\}\sim 0 &  \quad\{P\,,\,Q\} \sim 0 & 
\end{array}
\eeq

\subsection{Gauge invariance of the momentum and angular momentum}
We will now establish gauge invariance of the momentum and total angular momentum under the set of allowed internal gauge transformations. The previous calculation demonstrated gauge invariance of the total momentum and angular momentum under $g\sim o(\rho^{-1})$ basis transformations. It remains to demonstrate invariance under gauge transformations generated by $g\sim {}^0g+{}^1g/\rho+o(\rho^{-1})$, subject to the condition that ${}^0g$ is global, and ${}^1g$ is parity even. To this end, suppose we have two bases $e$ and $e_*$ related by a gauge transformation: $e_{*}=g e g^{-1}$. We need to establish the equivalence of $P=P(\bar{K}, e)$ and $L+S=L(\bar{K}, e)+S({}^0\lambda, e)$ with $P_{*}=P(\bar{K}, e_{*})$ and $L_{*}+S_{*}=L(\bar{K},e_*)+S({}^0 \lambda_*, e_*)$. We will demonstrate this under the assumption that $g$ is infinitesimal. We first note that the relation between ${}^0\lambda_*$ and ${}^0 \lambda$ is fixed by the requirement that $\bar{K}$ and ${}^0\lambda_*$ is a rigid isometry for ${}^0 e_{*}=e+[{}^0\chi, {}^0e]$ for $g\approx 1+{}^0\chi+{}^1\chi/\rho$. This yields the condition
\beqa
{}^0 \lambda_{*}={}^0\lambda+[{}^0\chi,{}^0 \lambda]\,.
\eeqa
The total expression then becomes
\beqa
L_{*}+S_{*}&=&-\frac{1}{k}\int_{\Sigma}\star\iota_{\bar{K}}(e_*\, e_*)\,\omega_{*}-\frac{1}{k}\int_{\Sigma}\star {}^0\lambda_{*} \,e_*\,e_* \nn \\
&=& L+S +\frac{1}{k}\int_{\Sigma}\rho^{-1}\,\star \iota_{\bar{K}}(e\,e)\,d{}^1\chi -\int_{\Sigma}\rho^{-1}\,\star [{}^0 \lambda , {}^1 \chi] e\,e
\eeqa
where we have used the constancy of ${}^0\chi$. Due to the parity conditions on ${}^1\chi$, the second term of the right hand side of the last line above vanishes. If $\bar{K}={}^0\bar{K}$ is a translation Killing vector the remaining term only involves ${}^0 e$ since in this case $\bar{K}\sim \rho^0$. In this case we have
\beqa
\frac{1}{k}\int_{\Sigma}\rho^{-1}\,\star \iota_{{}^0\bar{K}}(e\,e)\,d{}^1\chi&=&\frac{1}{k}\int_{\Sigma}\rho^{-1}\,\star \iota_{{}^0\bar{K}}({}^0e\,{}^0e)\,d{}^1\chi \nn\\
&=& \frac{1}{k}\int_{\Sigma}\star (\mathcal{L}_{{}^0\bar{K}} {}^0 e\,{}^0 e) \, {}^1\chi \nn\\
&=& -\frac{1}{k}\int_{\Sigma}\star [{}^1\chi, {}^0\lambda] {}^0e\,{}^0 e\nn\\
&=& 0
\eeqa
where the last line again follows by parity arguments. In the case that $\bar{K}={}^{-1}\bar{K}\sim \rho^{1} $ is a boost or rotational Killing vector, we have 
\beqa
\frac{1}{k}\int_{\Sigma}\rho^{-1}\,\star \iota_{{}^{-1}\bar{K}}(e\,e)\,d{}^1\chi&=&\frac{1}{k}\int_{\Sigma}\rho^{-1}\,2\,\star \iota_{{}^{-1}\bar{K}}({}^0e\,{}^1e)\,d{}^1\chi \,.
\eeqa
With the form of ${}^1 e$ given by (\ref{e1}), some calculation reveals that the above expression integrates to zero. Similarly, one can show that the linear momentum is invariant. Thus, combined with the invariance of $J_{tot}$ under ${}^2 \chi$ demonstrated in the previous section, we have
\beq
L_{*}=L\quad \quad S_{*}=S \quad \quad P_{*}=P\,,
\eeq
demonstrating the invariance of our expressions under the set of allowed internal gauge transformations.

\section{Summary and Conclusions}
In this paper we have argued that in the presence of spinorial matter sources, the ordinary angular momentum must be supplemented with an additional term corresponding to the internal spin angular momentum of the spacetime. The key algebraic signature of the existence of an internal spin structure can be seen form the structure of the isometry group, which is given by the isometry group of the flat tetrad, $\mathfrak{G}({}^0e)$ isomorphic to $Spin(3,1)\otimes(SO(3,1)\ltimes \mathbb{R}^{3,1})$, together with the requirement that the $SO(3,1)$ and $Spin(3,1)$ subgroups are ``locked" in order to preserve the flat tetrad ${}^o e$. This group structure, which we refer to as the spin-enlarged Poincar\'{e} algebra, is the basis for ordinary (spin+orbital) angular momentum in standard flat space quantum field theory. In the context of first order Einstein-Cartan theory, in order to arrive at a notion of conserved total angular momentum including internal spin, the isometry group of the spacetime must asymptotically reduce to the spin-enlarged Poincar\'{e} algebra. This means that the internal structure group of Einstein-Cartan gravity must at some level descend to the spin-enlarged Poincar\'{e} algebra so that $Spin(3,1)\rtimes Diff_4 \rightarrow Spin(3,1)\otimes(SO(3,1)\ltimes \mathbb{R}^{3,1})$ with the ``locking" condition on the Lorentz subgroups. This should now be the isometry group at asymptotic infinity, and the generators of these transformations on the phase space of asymptotically flat spacetimes should be related to the energy-momentum, orbital angular momentum, and spin-angular momentum. 

By assuming the existence of a global spin-enlarged isometry, we were able to demonstrate the relation between the Noether charges of the Dirac field, the asymptotic ADM-like boundary integral, and the Komar integral. Essential in the derivation was the inclusion of the additional term $S(\lambda)=-\frac{1}{k}\int\star\lambda\,e\,e$ to the total angular momentum. Without this term, the integrals would fail to pick up the internal spin of the matter fields in the bulk, and the integral identities would not hold.

By an explicit analysis of the asymptotic phase space, made possible by recent results on the finiteness of first-order gravity without infinite counter-terms, we demonstrated the finiteness of the integral expressions and the existence of a spin-enlarged Poincar\'{e} symmetry provided that certain restrictions are placed on the generators of the symmetry at asymptotic infinity. These restrictions ensure that at lowest order, the generators ${\bar{K}}^{\{\h{I}\h{J}\}}$ and $\lambda^{\{\h{I}\h{J}\}}$ do in fact generate the spin-enlarged Poincar\'{e} symmetry at asymptotic infinity, and at higher orders they generate the ``pure gauge" transformations of the bulk. As expected, under the Poisson bracket the resulting generating functionals do form a representation of the  the Lie algebra of $\mathfrak{G}({}^o e)$ together with the locking condition on the Lorentz subgroups. Given these algebraic properties, and the demonstrated relation among the functional generators, the Noether charge, and the Komar integral, we can safely identify the new term $L^{\h{I}\h{J}}+S^{\h{I}\h{J}}$ with the total relativistic angular momentum of the asymptotically flat spacetime. 

\section*{Acknowledgments}
We would like to thank Tomas Liko and Abhay Ashtekar for comments and discussions. AR would also like to thank Chris Beetle and Warner Miller for discussions during his visit to FAU. This research was supported in part by NSF grant PHY0854743, The George A. and Margaret M. Downsbrough Endowment and the Eberly research funds of Penn State. DS was also supported by a Frymoyer Fellowship.

\bibliography{MasterBibDesk}

\end{document}